\documentstyle[prl,aps,epsf]{revtex}
\begin{document}
\twocolumn[\hsize\textwidth\columnwidth\hsize\csname %
@twocolumnfalse\endcsname

\draft
\title{Charge ordering and long-range interactions in layered transition
metal oxides}
\author{Branko P.\ Stojkovi\'c$^{1}$, Z.\ G.\ Yu$^{1}$,
A.\ R.\ Bishop$^{1}$, A.\ H.\ Castro Neto$^{2}$ and
Niels Gr{\o}nbech-Jensen$^{1}$}
\address{$^{1}$Theoretical Division and Center for Nonlinear Studies, Los
Alamos
National Laboratory, Los Alamos, New Mexico 87545}
\address{$^{2}$Department of Physics, University of California, Riverside, CA
92521}
\date{\today}
\maketitle
\begin{abstract}
We study the competition between long-range and short-range
interactions among holes within the spin density wave 
picture of layered transition metal oxides. We focus on the problem
of charge ordering
and the charge phase diagram.
We show that the main interactions are the long-range
Coulomb interaction and a dipolar short-range interaction generated
by the short-range
antiferromagnetic fluctuations.
We find four different phases depending on the strength of the dipolar
interaction and the density of holes: Wigner crystal, diagonal stripes,
horizontal-vertical stripes (loops) and a glassy-clumped phase.
We discuss the effect of temperature, disorder and lattice effects
on these phases.
\end{abstract}
\phantom{.}

]

\narrowtext \pagebreak

Recently there has been much interest in the charge ordering and
domain wall formation at
mesoscopic scales in doped transition metal oxides \cite{review}.
A popular example of such orderings
are stripes, i.e., linear arrays of holes separated by an
antiferromagnetically (AF) ordered background. The formation
of domain walls and stripes has been discussed
in terms of the proximity to phase separation \cite{ps}.
Macroscopic phase separation has been observed in
La$_{2}$CuO$_{4+\delta}$\cite{hammel}, and stripes have been observed
experimentally in La$_{2-x}$Sr$_x$NiO$_{4+y}$ in
many different experiments including direct high-resolution
electron diffraction \cite{nickel}. Magnetic
susceptibility measurements\cite{suscep}, nuclear quadrupole resonance
\cite{nqr}
and muon spin resonance \cite{msr}
indicate formation of domains in La$_{2-x}$Sr$_x$CuO$_{4}$.
Stripes have also been seen in La$_{1-x}$Ca$_x$MnO$_3$
for specific commensurate values of doping \cite{stripemagneto}.
A direct evidence for stripe formation was observed
in neutron scattering in
La$_{1.6-x}$Nd$_{0.4}$Sr$_x$CuO$_{4}$ \cite{tranquada}.
Moreover, recent neutron scattering experiments
are not inconsistent with stripe phases in other high temperature
superconductors such as YBa$_2$Cu$_3$O$_{7-\delta}$ \cite{agree,aeppli}.

On the theoretical side, stripes have been proposed as a result of
a competition between short-range attractive interaction between holes
from the breaking of AF bonds and the long-range
Coulomb interaction \cite{emery1}. Indirect support for
this picture has been given in terms of the mapping of the problem into
effective spin models \cite{emery2}. Striped phases have been obtained
within mean field approaches to the short-range Hubbard or $t-J$ models
which are only able to generate insulating states
\cite{hartree-fock}. Numerical methods
in these models have not
been able to confirm this picture except for recent Density
Matrix Renormalization Group simulations (DMRG) \cite{white}.

In this paper we present a numerical approach to the problem of holes
moving in an AF insulator in the presence of long-range
Coulomb forces.  The ability to handle long-range Coulomb
interactions at finite density has been enhanced recently in the area
of molecular physics: assuming a computational cell of arbitrary
geometry and cyclic boundary conditions it is
possible to sum interactions of particles with all of their images
residing in cells obtained by translation from the original
computational cell\cite{lekner,niels}.
On making integral transformations, Coulomb
interactions are computed by summing over fast-convergent Bessel
functions with great accuracy.

\begin{figure}
\vskip -10pt
\centerline{
\epsfxsize=6.0cm \epsfbox{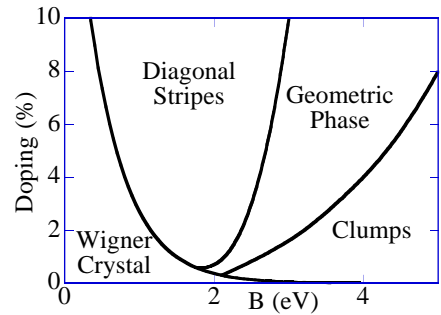}}
\vskip 5pt
\caption{Phase diagram as a function of the hole density and the strength
of the dipolar interaction, $B$, for $A=0$.}
\label{phdiag}
\end{figure}

Using Monte Carlo (MC) and
molecular dynamics (MD) methods, we systematically study
the interplay between long-range Coulomb interaction and short-range
AF interactions of dipolar nature
which we take to have both
isotropic and anisotropic components (depending on the lattice structure).
Our main result is summarized in the phase diagram
of Fig.\ref{phdiag}. In the absence of disorder
we find four phases depending on the density
of holes and the characteristic AF energy scales:
a Wigner crystal, diagonal stripes,
horizontal-vertical stripes (loops)
and a glassy-clumped phase. The  order parameter for charge ordering
is the Fourier transform of the hole density:
\begin{eqnarray}
\rho({\bf q}) = \frac{1}{N} \sum_{i=1}^N e^{i {\bf q} \cdot {\bf r}_i} \, ,
\label{rho}
\end{eqnarray}
where ${\bf r}_i$ is the position of the $i^{th}$ hole and
$N$ is the total number of holes. A peak in $\rho({\bf q})$
at some wave-vector ${\bf q}={\bf K}$ indicates
ordering.

Our starting point is the spin density wave (SDW)
 picture of the layered transition
metal oxides which has been very successful in describing the
insulating AF phase of these systems
\cite{slater,bob}. In this picture the electrons move with hopping
energy $t$ in the
self-consistent staggered field of its spin. Because the
translational symmetry of the system
is broken, the electronic band is split
into upper and lower Hubbard bands \cite{mott}. These
are separated by the Mott-Hubbard gap, $\Delta$, and at half
filling the lower band is filled and the upper one is empty.
This picture is consistent with the angle resolved photoemission data
in the layered AF insulator Sr$_2$CuO$_2$Cl$_2$ \cite{photo}.
By doping the system with holes with planar density $\sigma_s$
and at low temperatures, $T$
($k_B T << \Delta$), we focus entirely on the lower
band which has a maximum at ${\bf k}={\bf Q}/2 = (\pm 1,\pm 1)\pi/(2 a)$,
where $a$ is the lattice spacing.
It can be shown that the holes interact via two different mechanisms:
a short-range attractive force due to AF bond
breaking and a long-range dipolar interaction due to the
distortion of the AF background \cite{bob,ss}.
It was shown that this dipolar interaction gives rise to spiral distortions
of the AF background \cite{ss,david}. The dipole
moment associated with each hole is due to the virtual
hopping of holes between neighboring sites and scales with the
AF magnetic energy. The dipolar interaction between two holes
with dipole moments ${\bf d}_{1,2}$
at distance ${\bf r}$ apart has the form:
\begin{eqnarray}
U_{dip}=\frac{1}{r^2} \left[({\bf d}_1 \cdot {\bf d}_2)- \frac{2}{r^2}
({\bf d}_1 \cdot {\bf r}) ({\bf d}_2 \cdot {\bf r}) \right] \, ,
\label{dipole}
\end{eqnarray}
which is rotationally invariant. It is also possible to show
using Ward identities that the spin part of the
problem can be described by a two dimensional (2D) non-linear $\sigma$ model in
the long wavelength limit \cite{fradkin}. At finite $T$
the system is magnetically disordered and characterized by a finite magnetic
correlation length, $\xi$ \cite{chak}. Thus, at finite
T the dipolar interaction between the holes,
mediated by the antiferromagnet, is actually short-ranged. However, besides
the AF interactions the holes also feel the long-range
Coulomb interaction.
This is clear if we consider that $r_s=r_0/a_0$
(where $r_0$ is the mean inter-particle distance and $a_0$ is the
Bohr radius) is very large in the underdoped systems ($r_s \approx 8$).
Thus, the interaction energy between the holes, which behaves like
$e^2/(a_0 r_s)$, is certainly more important than
the kinetic term ($\approx e^2/(a_0 r^2_s)$) at low densities.
This implies that the interaction terms should
be treated first and the kinetic energy as a perturbation.
Finally, each hole carries a spin degree of freedom as well, but it
is possible to show that the overall spin energy is minimized in
the spin anti-symmetric channel, as we assume here.
Thus, in our approach, we are left with only the charge channel and
the interaction between two holes, $1$ and $2$, has the form
(see Eq.\ (\ref{dipole}))
\begin{eqnarray}
V({\bf r})= \frac{q^2}{r}
-A \, e^{-r/a}-B \cos (2\theta-\phi_1-\phi_2){e^{-r/\xi}} 
,\label{total}
\end{eqnarray}
where $q$ is the hole charge, $\theta$ is the angle made between
${\bf r}$ and a fixed axis and $\phi_{1,2}$ are the
angles of the dipoles
relative to the same fixed axis. $A$ is the strength of the short-range
bond-breaking interaction and $B$ is the strength of the dipolar interaction,
which we will assume to be independent variables. The magnetic
correlation length $\xi$ is obtained from neutron scattering
measurements \cite{birgeneau}.
It is also worth mentioning that the $1/2$-filled Landau
level problem has been mapped into an interacting 2D dipole gas
\cite{duncan}. More recently it was shown that the same
type of description is possible for a 2D electron gas even in
the absence of a magnetic field \cite{dhlee}.

In general, the many-body problem of holes in an AF
background is extremely complicated, involving many-particle
interaction terms. However, at low densities it is
reasonable to assume that the interaction of any two holes is weakly
perturbed by other holes, and the total potential energy can be
expressed in terms of two-particle energies.
Therefore, in our numerical calculations we study the physics of $N$ holes
interacting via $V({\bf r})$ as given in (\ref{total}).
We assume a rectangular
computational box of size $L_{x}\times L_y$ with $L_x$, $L_y$ up to
100 unit cells in a CuO$_2$ plane.
At the beginning of each simulation
we place the holes at random and assign to each hole a dipole
moment of constant size, but random direction.  We find a minimum of
the total potential in this system using three different methods: MC
method, Langevin MD and a hybrid MC-MD method\cite{janez}.  All three
methods yield essentially the same results. Since the
system exhibits several phases (see Fig.\ \ref{phdiag}) for some
values of the input parameters, its ground state is not always well defined
and may, in fact, depend on the initial and boundary
conditions. Hence, in order to rapidly reach a hole configuration with
the lowest global minimum energy we perform simulated annealing from
high temperatures.

\begin{figure}
\vskip 10pt
\centerline{
\epsfxsize=9.0cm \epsfbox{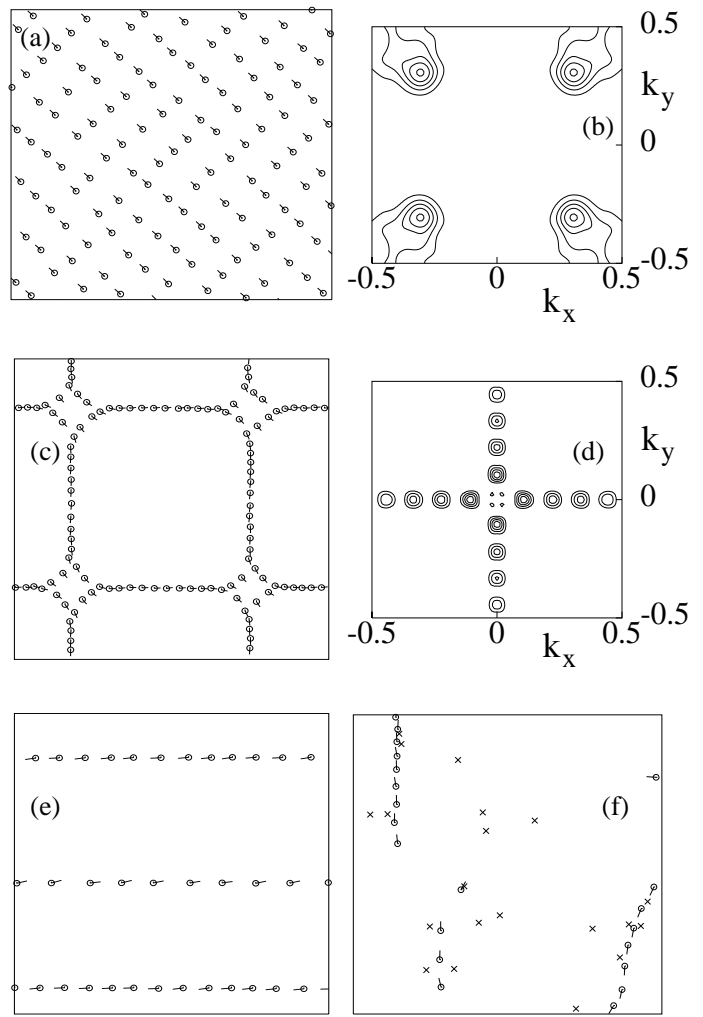}}
\vskip 5pt
\caption{Geometric phases resulting from the competition of dipolar
  and Coulomb interactions: panels (a) and (c) show holes (open
  circles) with their dipole orientations, in a small section of the
  computational box, for the ferro-dipolar and stripe phases,
  respectively; panels (b) and (d) show contour plots of the hole
  density in momentum space (see Eq. (1)) for the phases shown in
  panels (a) and (c). Panel (e) shows the stripe phase obtained with
  dipole anisotropy of 0.8, and panel (f) hole positions in the
  presence of impurities (crosses).}
\label{panels}
\end{figure}

For $B=0$ we find the Wigner crystal with small distortions to be
the state of lowest energy, as expected \cite{wigner}.
The small distortion of the crystal structure is due to the
periodicity, which introduces a small spatial anisotropy into the system
due to the rectangular shape of the computational box.  Increasing $A$
while retaining $B=0$ reduces the lattice constant of the Wigner
crystal until a critical value is reached where holes group together.
For $A=0$ and finite $B$ the situation is quite different.  At small
$B$ and larger densities the Wigner crystal is unstable and a new
phase with diagonal stripes is formed.  This phase is characterized by
ferro-dipolar order (see Fig.\ \ref{panels}a).  The situation here is
very similar to that observed in La$_{2-x}$Sr$_x$NiO$_{4+y}$
\cite{nickel}.  As shown in Fig.\ \ref{panels}c, at larger values of
$B$ a line stripe is formed, which, with increasing density tends to
close into loops, forming a checkerboard pattern.
Importantly, the loop formation is accompanied by dipole orientation
along the straight portion of a loop with gradual rotation by $\pi/2$
at each corner \cite{vertex}. Due to the rotation of dipoles at
corners the loops interact, and eventually form the checkerboard
pattern \cite{ferro-vertex}.
The size of the inter-hole distance within
a line is determined by the ratio of $B$ and the Coulomb energy; the
loop sizes are determined by the hole density alone.  These results
appear to be consistent with the DMRG solution of the t-J model
\cite{white}.  If $B$ is increased further the dipolar interaction
becomes dominant over the average Coulomb interaction; the 
well-defined pattern disappears and one observes star shaped clumps of
holes, which can, at sufficiently high density, form another geometric
structure (e.g., a Wigner crystal of clumps).
We remark that in all phases a non-vanishing value of $A$
leads to a decrease in the effective value of $B$ at which the
transitions occur (Fig.\ \ref{phdiag}); the isotropic term $A$ alone
{\em never} produces any non-trivial geometric phase (e.g., stripes),
even with inclusion of lattice effects.  We find that the transition
between the ferro-dipolar and the stripe phase is first order, while
other transitions appear to be of second order \cite{unpu}. The stripe
tension of the hole patterns will be quantified elsewhere \cite{unpu}.

In the cases presented above we have assumed uniform dipolar
interaction.  It is well known that there are orthorhombic and
tetragonal distortions in practically all transition metal oxides.
In particular static stripe formation has only been observed in the
low temperature tetragonal phase of
La$_{1.6-x}$Nd$_{0.4}$Sr$_x$CuO$_{4}$ \cite{tranquada}.  In order to
study the influence of the anisotropy we assume that the dipole sizes
along $x$ and $y$ directions have anisotropy $\alpha$ ($\alpha=1$
corresponds to the isotropic case). Figure
\ref{panels}c shows our solution for $\alpha=0.8$:
the symmetry is broken and a stripe superlattice is formed,
with a charge ordering vector ${\bf K} = (\pi/\ell) {\bf x}$, where
$\ell$ is the inter-stripe distance.  In the SDW model the Fourier
transform of the magnetization $S({\bf q}) = \langle S_z({\bf q})
\rangle$ is slaved to (\ref{rho}) such that a peak at ${\bf
  K}$ in (\ref{rho}) leads to a peak at ${\bf Q} \pm {\bf K}$ in
$S({\bf q})$ \cite{unpu}. Thus our results yield a neutron peak at
$(\pi/a \pm \pi/\ell,\pi/a)$.  Assuming twinning,
this would imply neutron peaks at $(\pi/a \pm \pi/\ell,\pi/a
\pm \pi/\ell)$ in agreement with experiment\cite{aeppli}.
The same is obtained in the checkerboard phase (see Figs.\
\ref{panels}c and \ref{panels}d).  If one
includes the kinetic energy\cite{unpu}, instead of
static stripe formation one would obtain dynamical stripes like those
believed to exist in La$_{2}$Sr$_x$CuO$_{4}$. In this case the
Fermi surface of the system should be modified by the superlattice
formation \cite{paco}.

Our results are somewhat sensitive to the applied boundary conditions:
first, the exact size of the checkerboards depends on its
commensuration with the computational box, which, in turn depends on
the density. On increasing of the size of the computational box, the
checkerboard pattern shown in Fig.\ \ref{panels}c acquires point or
line defects\cite{unpu}. This leads to the reduction in the higher
order peaks observed in Fig.\ \ref{panels}d with no change in their
wave-numbers. Second, in a finite system with appropriate charge
background, the holes do not form geometric phases, although
they still form stripes\cite{unpu}. However, in this case even a
very small anisotropy ($\alpha\sim 0.95$) again leads to stripe formation
as in Fig.\ \ref{panels}e \cite{unpu}.

We have also studied impurity effects (from defects or charged
counter-ions). For example, we place the same number of
impurities as holes, randomly in a plane a distance
$d=6 \AA$ above the plane to simulate the situation in, e.g., Sr doped
cuprates and consider the {\it unscreened} attractive Coulomb
interaction between impurity and hole. The charge pattern produced is
very sensitive to impurity doping (see Fig.\ \ref{panels}f). The
Wigner crystal becomes glassy with no obvious sign of charge ordering.
This happens because the attractive
Coulomb energy between impurities and holes scales like
$e^2/d$ while the average inter-hole Coulomb energy behaves like
$e^2 \sqrt{\sigma_s}$. Thus when $\sigma_s<1/d^2$ the holes are pinned
by impurities. Most strikingly, all other phases are unstable towards {\it
finite} stripe
formation. The loops and diagonal stripes tend to
deform to pass very close to the impurities in order to
maximize the attractive energy.  However, the dipole interaction is
sufficient to retain the main orientation. This leads us to
conjecture that with the addition of the kinetic energy the holes can
move in string segments in an
orientation given basically by the phase diagram of the clean system.
These string segments are kept together by the dipolar interaction (i.e.,
string tension).  The stripe motion would then be caused by
mesoscopic thermal or quantum tunneling of the finite strings between
the minima of the overall potential. This would lead to
non-linear effects in the low temperature field dependent conductivity
\cite{bardeen} and unusual $T$ dependence of the
conductivity \cite{unpu}.  We have also performed simulations in the
presence of a realistic underlying periodic lattice and have found
that this creates slight distortions in the phases, pinning
loops more strongly\cite{unpu}.  In
particular, the peaks in $\rho({\bf q})$ sharpen in some of these
phases.  Finally, at finite $T$ melting of the phases
occurs because of the small energy scales and large entropy in long-range
Coulomb tails.

In summary, using a novel numerical technique, we have studied the
competition between long-range and short-range interactions and its
impact on hole ordering in layered transition metal oxides.
Employing the SDW picture of these systems, we have studied the
short-range attractive force and the dipolar force generated by the
short-range AF fluctuations together with long-range Coulomb
forces for a 2D layer.  We have found a rich phase diagram for the
clean system which includes a Wigner solid, stripes, loops and a
glassy phase.  This phase diagram is consistent
with several different experimental measurements.  We have also
found this system to
be rather sensitive to the presence of charged impurities. However, the
stripe phases survive as finite stripe segments which we believe is
key to understanding these systems. Finally we have also
investigated the effects of a periodic lattice, which leads to minor
distortions of the various phases. These results are of great
importance for the understanding of the phase diagram of layered
transition metal oxides.

We gratefully acknowledge valuable discussions with A.\ Balatsky,
A.~Chernishev, J.\ Gubernatis, C.~Hammel, D.\ Pines, J.\ Schmalian,
S.~White and J.~Zaanen.  A.~H.~C.~N. acknowledges support from the
Alfred P.~Sloan foundation.  Work at Los Alamos was supported by the
U.\ S.\ Department of Energy.  Work at the University of California,
Riverside, was partially supported by a Los Alamos CULAR project.

\begin{references}
\bibitem{ferro-vertex} The simulations show the presence of
  topological defects in the phase of the dipole moments, i.e.,
  the total phase in a loop may be shifted by $2\pi$.
\bibitem{unpu} Details of this calculation will be given elsewhere.
\end{references}
\end{document}